\documentclass[prl,twocolumn]{revtex4}
\usepackage{amsmath}
\usepackage{amssymb}
\usepackage{bm}
\usepackage{epsfig}
\usepackage{graphicx}

\begin{document}

\title{Interplay of Rashba and Dresselhaus spin splittings \\ in
2D weak localization}

\author{M.M.~Glazov and L.E.~Golub}

\affiliation{A.F. Ioffe Physico-Technical Institute, Russian
Academy of Sciences, 194021 St.~Petersburg, Russia}

\begin{abstract}
{The effects of structural (Rashba) and bulk (Dresselhaus)
spin-orbit interaction terms on the low-field magnetoresistance
are investigated in high-quality two dimensional systems. The weak
localization theory accounting for both of these terms valid in
the whole range of magnetic fields is proposed. The suppresion of
antilocalization correction as Rashba and Dresselhaus terms
strengths approach each other is demonstrated. The effect of cubic
in the wavevector spin-splitting term is analyzed.}
\end{abstract}

\maketitle

\section{Introduction}

Spin properties of carriers in semiconductor heterostructures
attract now great attention due to spintronics proposals. From the
point of view of fundamental physics as well as of possible
applications the control of spin-orbit splitting of electronic
energy spectrum in two-dimensional systems is one of the most
important aspects. There are two contributions to the spin-orbit
splitting in two-dimensional semiconductor structures: the Rashba
and the Dresselhaus terms caused by structure and bulk inversion
asymmetry respectively. Their magnitudes can be measured in
various optical and transport experiments. In this work we
investigate the effect of these spin splittings on low-field
magnetoresistance caused by weak localization.

As the nature of Dresselhaus and Rashba terms is different they
possess different symmetry and they are not additive in various
kinetic phenomena. The weak antilocalization theory developed in
Ref.\cite{Glazov01} was limited to extremely weak magnetic fields
where the electron motion can be considered as diffusive and to
small values of spin-splittings. The theory valid in the whole
range of the magnetic fields was developed in the
Ref.\cite{Glazov03} for the case of only one term being relevant.
However, the experiment evidences the comparable magnitudes of
both contributions, see e.g. Ref.\cite{Glazov02}. An attempt to
account simultaneously both Dresselhaus and Rashba terms was
carried out in Ref.\cite{Glazov02} but their theory was limited by
rather high magnetic fields.

Here we present a general theory which allows to compute quantum
conductivity corrections in the whole range of classically weak
magnetic fields and for arbitrary values of Dresselhaus and Rashba
spin-orbit splittings. We demonstrate that, for equal magnitudes
of these contributions, the weak-localization conductivity
correction exactly equals to that in the absence of spin-orbit
interaction. In such a case even small cubic in wavevector
spin-splitting modifies strongly of magnetoresistance leading to
antilocalization minimum.

\section{Theory}

In what follows we concentrate on zince-blende lattice-based
quantum wells grown along $[001]$ direction. The spin-orbit
interaction is described by the following Hamiltonian
\begin{equation}\label{hso}
    H_{SO}(\bm{k}) = \hbar  \: \bm{\sigma} \cdot
    [ \bm \Omega_R(\bm k) + \bm \Omega_D(\bm k)],
\end{equation}
where $\bm{k}$ is the electron wave vector, $\bm{\sigma}$ is the
vector of Pauli matrices. $\bm{\Omega}_{R(D)}$ characterizes the
spin precession frequency due to Rashba (Dresselhaus) term which
explicitly read: $\bm \Omega_D(\bm k) = \Omega_D (\cos{\chi}, -
\sin{\chi})$ and $\bm \Omega_R(\bm k) = \Omega_R (\sin{\chi}, -
\cos{\chi})$. Here $\chi$ is the angle between $\bm k$ and $[100]$
axis and $\Omega_D$, $\Omega_R$ are the coupling strengths.

In accordance with Eq.~(\ref{hso}) the electron motion is
accompanied with spin rotation yielding the appearance of
spin-dependent phase in the Green's function~\cite{Glazov03}
\begin{equation}\label{Greens}
G^{R,A}(\bm r, \bm r') = G_0^{R,A} (R) \exp{[\mathrm i \varphi(\bm
r,\bm r') - \mathrm i \bm \sigma \cdot \bm\omega(\bm R)]},
\end{equation}
where $\bm R = \bm r - \bm r'$, $G_0^{R,A}$ are retarded ($R$) and
advanced ($A$) Green's function for electron's propagation in the
short-range random potential without magnetic field $B=0$ and spin
splitting $\bm \Omega=0$, $\varphi(\bm r,\bm r') =
(x+x')(y'-y)/2l_B^2$ with $l_B = \sqrt{\hbar/eB}$ being the
magnetic length and vector $\bm\omega(\bm R) = \bm{\Omega}_R(k_F
l^{-1}\bm R) + \bm{\Omega}_D(k_F l^{-1}\bm R)$, $k_F$ is the Fermi
wavevector, $l$ is the mean free path.

Following the general theory of quantum-conductivity
corrections~\cite{Glazov04} we solve the equation for the
interference magnitude Cooperon $\mathcal C(\bm r, \bm r')$
\begin{equation}\label{C_Eq}
{\cal C}(\bm{r},\bm{r}') = {\hbar^3 \over m \tau} \mathcal
P(\bm{r},\bm{r}') +
    \int d \bm{r}_1 \mathcal P(\bm{r},\bm{r}_1) {\cal
    C}(\bm{r}_1,\bm{r}'),
\end{equation}
where $m$ is electron effective mass, $\tau$ is the scattering
time, $\mathcal P_{\alpha\gamma,\beta\delta}(\bm r, \bm r')=
\hbar^3/m\tau \:G_{\alpha\beta}^{R}(\bm r, \bm
r')G_{\gamma\delta}^{A}(\bm r, \bm r')$, expanding the quantities
entering in Eq. (\ref{C_Eq}) in the series over the wavefunctions
of spinless particle with charge $2e$. In the general case where
both Rashba and Dresselhaus terms are present in the spin-orbit
Hamiltonian, the isotropy of the energy spectum in the system is
removed and all the `Landau levels' of $2e$ particle become
intermixed as opposed to the case of one term being relevant. In
the latter situation the total angular momentum is conserved
allowing to separate full Cooperon matrix into finite
blocks~\cite{Glazov03}. Finally, as Cooperon is known the
computation of conductivity reduces to the simple linear-algebra
problem~\cite{Glazov05}.

\section{Results and Discussion}

\begin{figure}[t]
\hskip-1.5 cm\epsfxsize=0.85\linewidth \leavevmode
{\epsfbox{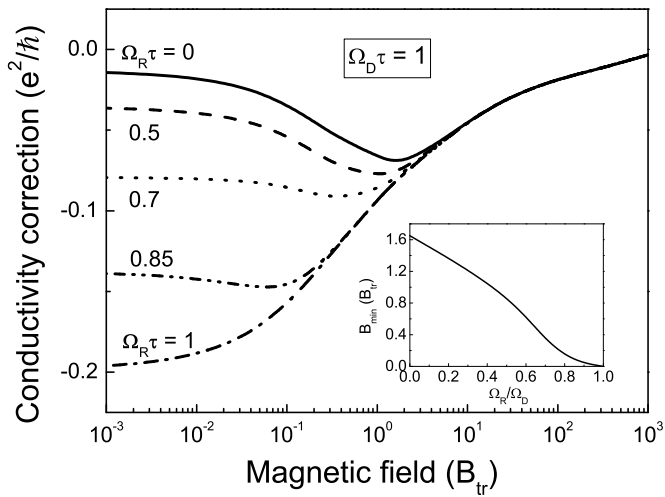}} \caption{Magnetoconductivity computed for
the constant Dresselhaus term $\Omega_{D}\tau = 1$ and different
values of Rashba term $\Omega_{R}\tau$. The phase-breaking time
$\tau_\phi = 100\tau$. The inset shows the minimum position vs.
$\Omega_R/\Omega_D$.}\label{k-linear}
\end{figure}

Figure \ref{k-linear} presents the quantum conductivity correction
plotted vs. $B/B_{tr}$ ($B_{tr} = \hbar/2el^2$) for different
relation of Rashba and Dresselhaus constants. The value of
Dresselhaus term is fixed to be $\Omega_{D}\tau = 1$ for all
curves in the figure while Rashba term took different values
$\Omega_R\tau$.

The curve corresponding to $\Omega_R=\Omega_D$ exactly coincides
with the result for spinless particles~\cite{Glazov04}. The
magnitude of the quantum correction decreases with the magnetic
field as long trajectories (longer than $l_B$) are suppressed in
the magnetic field. In this case the spin precession axis is
directed along either $[1\bar 1 0]$ or $[110]$ for any wavevector
$\bm k$ and the spin rotation angle for closed trajectories is
zero. From the point of view of the quantum mechanics the
single-particle spectrum in this case consists of two independent
paraboloids which contribute additively to the
magnetoconductivity~\cite{Glazov06}.

With the decrease of the Rashba term, the spin rotation angle for
closed loops is no longer equals to zero and the
magnetoconductivity becomes non-monotonous. As for the given
trajectory the spin rotation angle will be the greater the greater
$|\Omega_R^2 - \Omega_D^2|\tau^2$, the decrease of the Rashba term
(at fixed Dresselhaus one) will manifest itself as an increase of
spin-orbit interaction~\cite{Glazov01}. The minimum of the
magnetoconductivity shifts to the smaller fields (see inset to
Fig.~\ref{k-linear}) and its depth decreases.

\section{Effect of $\bm k$-cubic term}

\begin{figure}[t]
\hskip-1.5 cm\epsfxsize=0.85\linewidth \leavevmode
\epsfbox{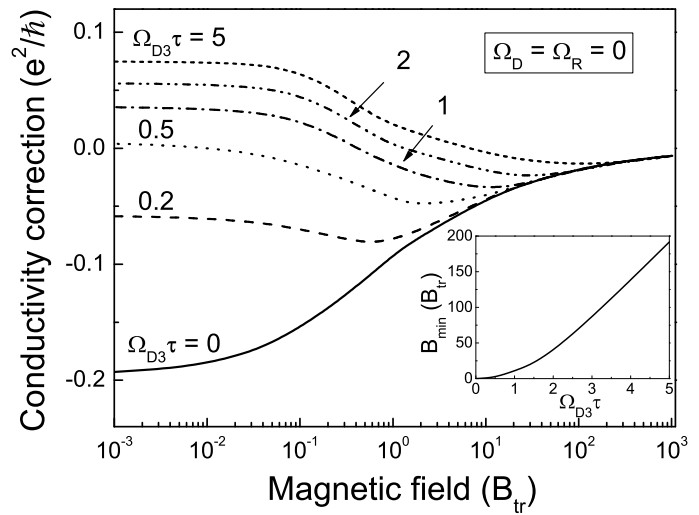} \caption[]{Conductivity correction computed
for case of zero linear-$\bm k$ contribution $\Omega_R = \Omega_D
=0$ and different values of $\bm k$-cubic term $\Omega_{D3}\tau$
and $\Omega_D = \Omega_R =0$. The inset demonstrates the
antilocalization minimum position vs.
$\Omega_{D3}\tau$.}\label{k-cubic}
\end{figure}

Besides the linear-$\bm k$ terms studied above the spin-orbit
interaction Eq. (\ref{hso}) contains also cubic in the wavevector
contribution originated from the bulk Dresselhaus term
\begin{equation}\label{k3}
\bm \Omega_3^D(\bm k) = \Omega_{D3} (\cos{3\chi}, \sin{3\chi}).
\end{equation}
The constant $\Omega_{D3}$ can be of the same order of magnitude
as $\Omega_{D}$ and $\Omega_R$~\cite{Glazov01,Glazov02} thus its
effect on quantum conductivity corrections may be important. Also,
the calculation of the zero-field correction shows that for
$\Omega_R/\Omega_D=1$ where linear in $\bm k$ terms compensate
each other the conductivity correction is determined to within the
$10\%$ by cubic contribution Eq. (\ref{k3}). Thus, we focus on the
case where only $\bm k$-cubic term is present in the effective
Hamiltonian and compute magnetoconductivity for the different
values of the $\Omega_{D3}\tau$ and $\Omega_D = \Omega_R =0$, see
Figure \ref{k-cubic}.

Qualitatively the behavior of the curves in Figure \ref{k-cubic}
concides with that found for case of linear in the wavevector spin
splitting~\cite{Glazov03}. In the high field case $B\ll
\max{[(\Omega_{D3}\tau)^2,1]}B_{tr}$ all curves have the same
asymptotics. As it can be seen from the figure inset the
magnetoconductivity minimum shifts to the higher field range with
an increase of $\Omega_{D3}\tau$. The minimum depth is a
non-monotonous function of the spin splitting: for the small spin
splitting values the increase of the splitting leads to an
increase of mininum depth while for higher splittings the behavior
is opposite.

\section{Conclusion}
We put forward a general theory of quantum conductivity
corrections valid in the whole range of classically weak magnetic
fields and taking into account the conduction band spin-splitting.
Both Rashba and Dresselhaus terms were considered. We have shown
that if Rashba and Dresselhaus terms are exactly equal the
weak-localization correction is the same as in the spin-less case.
The crossover between weak localization and weak antilocalization
is demonstrated as Rashba term value approaches to that of
Dresselhaus term. The effect of cubic in the wavevector part of
Dresselhaus term is comprehensively studied. The proposed theory
can be used in the analysis of the low-field magnetoresistance to
precisely extract the spin-splitting magnitudes from the transport
measurements.

This work is financially supported by
RFBR, Russian President grant for young scientists and by
``Dynasty'' Foundation
--- ICFPM.

\end{document}